\documentclass[fleqn,12pt,twoside]{article}
\usepackage[headings]{espcrc1}
\usepackage{epsfig}
\usepackage{float}
\usepackage{subfigure}

\newcommand{\eq}[1]{\begin{equation}#1\end{equation} }

\newcommand{\eqsa}[1]{\begin{eqnarray}#1\end{eqnarray} }
\newcommand{\ce}[1]{Eq.~(\ref{#1})}

\newcommand{\cf}[1]{{Fig.~\ref{#1}}}

\newcommand{\ep}{\varepsilon}

\newcommand{\ks}[1]{#1 \!\!\! \slash \, \!}

\newcommand{\ga}{\gamma^5}
\newcommand{\gmu}{\gamma^\mu}
\newcommand{\gnu}{\gamma^\nu}

\newcommand{\ie}{{\it i.e.}}
\newcommand{\eg}{{\it e.g.}}
\newcommand{\nn}{\nonumber}
\renewcommand{\cal}{\mathcal}
\newcommand{\etal}{{\it et al.}}

\newcommand{\beq}[1]{
\begin{equation}\label{#1}}
\newcommand{\eeq}{\end{equation}}
\newcommand{\bea}[1]{
\begin{eqnarray}\label{#1}}
\newcommand{\eea}{\end{eqnarray}}
\readRCS
$Id: espcrc1.tex,v 1.2 2004/02/24 11:22:11 spepping Exp $
\usepackage{graphicx}
\usepackage[figuresright]{rotating}

\newcommand{\AmS}{{\protect\the\textfont2
  A\kern-.1667em\lower.5ex\hbox{M}\kern-.125emS}}

\title{\center Backward DVCS and Proton to Photon Transition Distribution Amplitudes}

\author{J.P.~Lansberg\protect\footnote{Jean-Philippe.Lansberg@cpht.polytechnique.fr}
\address[CPHT]{CPHT\protect\footnote{Unit\'e mixte 7644 du CNRS},  \'Ecole Polytechnique , 91128 Palaiseau, France}\address[PTF]{Physique Th\'eorique Fondamentale, Universit\'e de Li\`ege,
17 All\'ee du 6 Ao\^ut, B\^at. B5, B-4000 Li\`ege-1, Belgium},
B.~Pire\addressmark[CPHT],
and L.~Szymanowski \addressmark[PTF]\address[SI]{Soltan Institute for Nuclear Studies, Warsaw, Poland}\address[LPT]{LPT\protect\footnote{Unit\'e mixte 8627 du CNRS }, Universit\'e Paris-Sud, 91405, Orsay, France}}

\runtitle{Proton to Photon TDAs}
\runauthor{J.P.~Lansberg \etal}

\begin{document}

% typeset front matter
\maketitle

\begin{abstract}
We analyse deeply-virtual Compton scattering on a proton target, $\gamma^* P \to P' \gamma $  
 in the backward region  and in the scaling regime. We define the transition distribution 
amplitudes  which describe the proton to photon transition. Model-independent predictions 
are given to test  this description, for current or planned experiments 
at JLab or by Hermes. 
\end{abstract}

\section{INTRODUCTION}

Deeply virtual Compton scattering (DVCS) at small momentum transfer $t$ has been the subject of a continuous
progress in recent years, both on the theoretical side with the understanding of factorisation 
properties which allow a consistent calculation of the amplitude in the framework of QCD, and
on the experimental side with the success of experiments at HERA and JLab. The  generalised
 parton distributions (GPDs) which describe the soft part of the scattering amplitude indeed contain 
 much information on the hadronic structure, which would remain hidden without this new
 opportunity~\cite{PS}. In Ref.~\cite{TDApigamma}, it has been advocated that the same virtual Compton scattering 
 reaction 
  \begin{equation}
  e P(p_1)  \to e' P'(p_2) \gamma(p_\gamma) 
  \label{bvcs}
 \end{equation}
 as well as electroproduction of meson ($\pi$, $\rho$, \dots)
  \begin{equation}
  e P(p_1)  \to e' P'(p_2) M(p_M) 
  \label{bepm}
  \end{equation}
 in the backward kinematics (namely small $u=(p_\gamma - p_1)^2$ or $u=(p_M -p_1)^2$) could be analysed in a slightly 
 modified framework, the amplitude being factorised 
(see~\cf{fig:fact-ampl} (a) and (b)) at leading twist as
\eq{
{\cal M} (Q^2,\xi,\Delta^2)\propto \int dx_i dy_j \,\Phi(y_j,Q^2) M_h(x_i,y_j,\xi) \,T(x_i,\xi,\Delta^2)\;,
\label{amp}
}
where $\Phi(y_j,Q^2)$ is  the proton distribution amplitude, $M_h$ is a perturbatively calculable 
 hard scattering amplitude and $T(x_i,\xi,\Delta^2)$ are transition distribution amplitudes (TDAs)
 defined as the matrix elements of light-cone operators between a proton and a photon state or
between a proton and a meson state.

The variable $x_i$ describes the fraction of light-cone momentum
carried by the quark $i$ off the initial proton, $y_j$ is the corresponding one for the 
quark $j$ entering the final state proton, $\Delta=p_\gamma-p_1$ and the skewness 
variable $\xi$ describes the loss of plus-momentum of the incident proton (see section 2 for 
more details on kinematics).

In the large angle regime (around 90 degrees), the large value of $-t=-(p_1-p_2)^2$ sets the perturbative
scale. In the small angle regime as well as for the backward regime,
it is the large virtuality $Q^2$ of the initial photon
which allows a perturbative expansion of a subprocess scattering amplitude. Of course in the backward regime, small 
$-u=-(p_\gamma - p_1)^2$ means large $-t$, 
and even $-t$ larger than at 90 degrees, but this does not introduce a new 
scale in the problem, exactly as for the forward DVCS case for which, $-t$ being small, $-u$ is very large.

\begin{figure}[H]
  \centering{
\subfigure[$\gamma^\star P \to P' \gamma$]{\includegraphics[height=4.5cm,clip=true]{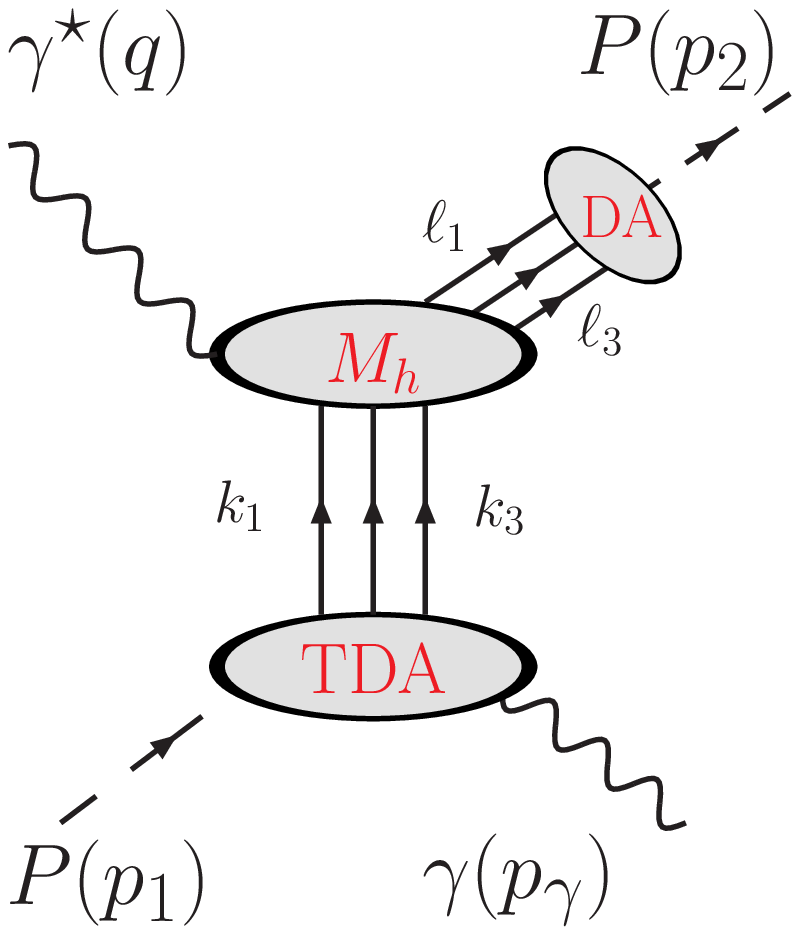}}~
\subfigure[$\gamma^\star P \to P M$]{\includegraphics[height=4.5cm,clip=true]{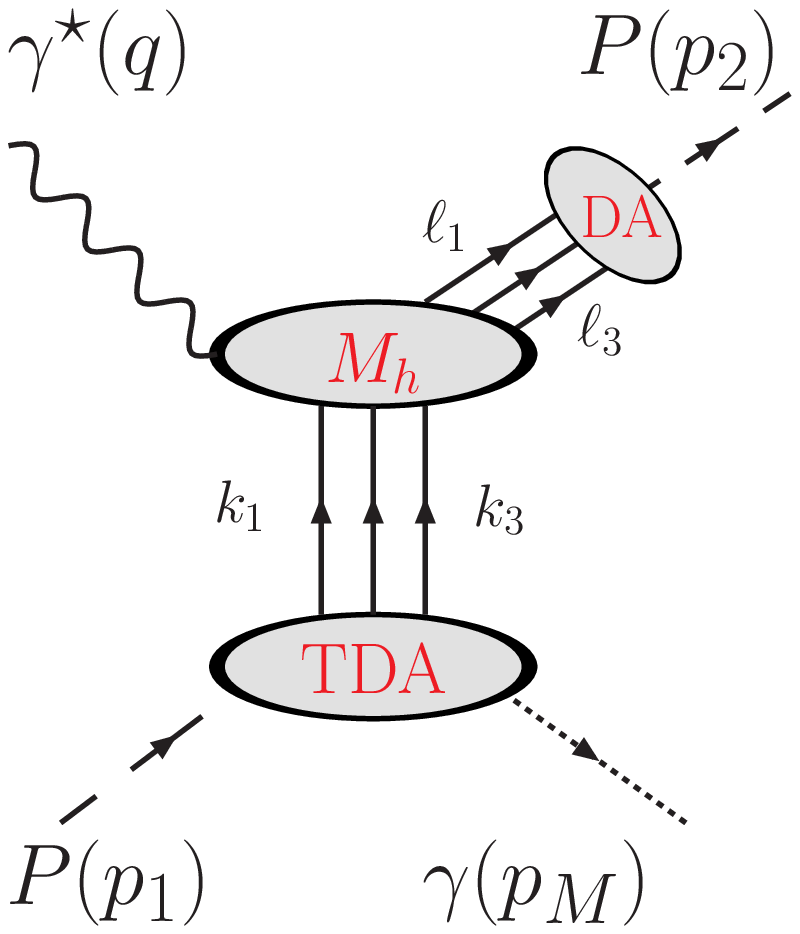}}
\subfigure[$\gamma^\star \gamma \to A \pi$]{\includegraphics[height=4.5cm,clip=true]{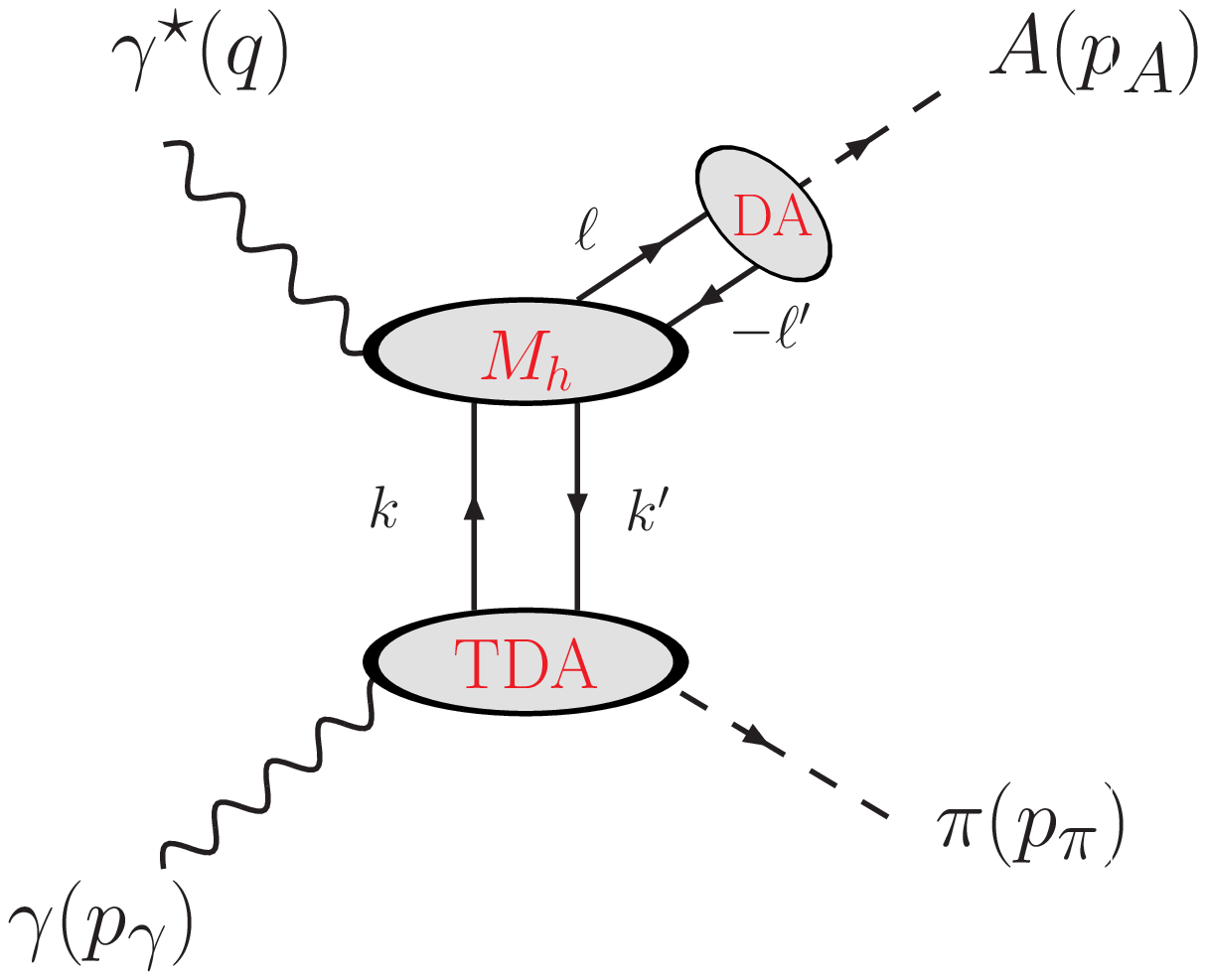}}~
}
\caption{ (a) Factorised amplitude for deeply-virtual Compton scattering on proton in the backward region;
(b) Factorised amplitude for meson electroproduction on proton in the backward region.
(c) Factorised amplitude for meson-pair ($A\pi$) production in $\gamma^\star\gamma$ collisions.}
\label{fig:fact-ampl}
\end{figure}

In Ref.~\cite{TDApiproton}, we have defined the leading-twist proton to pion
$P \to \pi$ transition distribution amplitudes from the Fourier transform\footnote{In the following,
we shall use the notation $\displaystyle {\cal F}\equiv (p.n)^3\int^{\infty}_{-\infty} dz_i e^{\Sigma_i x_i z_i p.n}$.} 
of the matrix element
\begin{equation}\label{eq:mat_el_p-pi}
   \langle \pi|\, \epsilon_{ijk} {q}^i_{\alpha}(z_{1}n)\, 
[z_1;z_0]\,{q}^j_{\beta}(z_{2} n)\, [z_2;z_0]\,
   {q}^k_{\gamma}(z_{3} n)\,[z_3;z_0] \,|P \rangle ,
 \end{equation}
  
The brackets $[z_i;z_0]$ in \ce{eq:mat_el_p-pi} account for the insertion of a 
path-ordered gluonic exponential 
along the straight line connecting an arbitrary initial point $z_0 n$ and a final one $z_i n$: 
\eqsa{
   [z_i;z_0] \equiv {\rm P\ exp\,}
   \biggl[ ig\int_0^1\!dt\, (z_i-z_0)n_\mu A^\mu(n[tz_i+(1-t)z_0])\biggr].
}
 which provide the QCD-gauge invariance for non-local operator and
equal unity in a light-like (axial) gauge. 

In a similar way, we shall define in section \ref{sec:ptogamma} the proton to photon TDAs 
from the Fourier transform of the matrix element
\begin{equation}
   \langle \gamma|\, \epsilon_{ijk} {q}^i_{\alpha}(z_{1} n)\, 
[z_1;z_0]\,{q}^j_{\beta}(z_{2} n)\, [z_2;z_0]\,
   {q}^k_{\gamma}(z_{3} n)\,[z_3;z_0] \,|P \rangle.
 \end{equation}  

In the simpler mesonic case, a perturbative limit has been 
obtained~\cite{Pire:2006ik}
for the $\rho$ to $\gamma^\star$ transition. For
$ \pi \to \gamma$ one, where there are only four leading-twist 
TDAs~\cite{TDApigamma} entering the parametrisation of the  matrix element
$   \langle \gamma|\, {\bar q}_{\alpha}(z_{1} n)\, 
[z_1;z_0]\,{ q}_{\beta}(z_{0} n) \,|\pi \rangle$,
we have recently shown~\cite{TDApigamma-appl} that experimental analysis of processes such
as $\gamma^\star \gamma \to \rho \pi$ and  $\gamma^\star \gamma
 \to \pi \pi$, see~\cf{fig:fact-ampl}
(c), involving 
these TDAs could be carried out,~\eg~the background from the Bremsstrahlung
is small if not absent and rates are sizable at present $e^+ e^-$ facilities.

\section{Kinematics}

The momenta of the processes $\gamma^* P \to P' \gamma $ are defined as shown in \cf{fig:fact-ampl} (a).
The $z$-axis is chosen along the initial nucleon momentum and the $x-z$ plane is identified 
with the collision plane. Then, we define the light-cone vectors $p$ and $n$ ($p^2$=$n^2$=0) 
such that $2~p.n=1$, as well as $P=\frac{1}{2} (p_1+p_\gamma)$, $\Delta=p_\gamma -p_1$ and its 
transverse component $\Delta_T$, which we choose to be along the $x$-axis. From those, we define 
$\xi$ in an usual way as $\xi=-\frac{\Delta.n}{2P.n}$.

We  can then express the momenta of the particles through their  
Sudakov decomposition:\footnote{
$\Delta_T^2<0$.}
\eqsa{\label{eq:decomp_moment}
\lefteqn{p_1= (1+\xi) p + \frac{M^2}{1+\xi}n, ~~~p_\gamma= (1-\xi) p -\frac{\Delta_T^2}{1-\xi}n+ \Delta_T,}\nn\\
\lefteqn{p_2= (2\xi-1) p + n [Q^2 +\frac{\Delta_T^2}{1-\xi}- \frac{M^2}{1+\xi}] - \Delta_T, ~~~q=-p + Q^2 n.}
}

Using the natural gauge choice $\ep.n=0$, the photon polarisation vector $\ep(p_\gamma)$ can 
be chosen to be either  a normalised vector along the $y$-axis, 
\eqsa{\ep_{T_1}=\ep_y ~~~\hbox{or} ~~~\ep_{T_2}=\frac{\Delta_T}{\sqrt{-\Delta_T^2}}+ 2\frac{\sqrt{-\Delta_T^2} }{1-\xi} n,}
which gives $\ep_{T_2}=\ep_x$ at $\Delta_T=0$.

In an arbitrary QED gauge, where $\ep'=\ep_T + \lambda p_\gamma$, we have at $\Delta_T=0$
\eqsa{
  \ep'_1=\lambda (1-\xi) p + \ep_y,~~\ep'_2&=\lambda (1-\xi) p+\ep_x.
}

Therefore one has, in any gauge and at $\Delta_T=0$,
\eqsa{\label{eq;pr-sc-ep}
\ep.p=0, ~~\ep.n=\lambda\frac{1-\xi}{2},~~\ep.\Delta_T=0.
}

\section{The Proton to Photon TDAs}\label{sec:ptogamma}

   The spinorial and Lorentz decomposition of the matrix element
will follow the same line as the one for  $P \to \pi$ TDA~\cite{TDApiproton} and for baryon 
DA~\cite{Braun:2000kw}. The fractions of plus momenta are labelled $x_{1}$, $x_{2}$ and $x_{3}$, 
and their supports are within $[-1+\xi, 1+\xi]$.  
Momentum conservation implies (we restrict to the case $\xi > 0$ ):
\begin{equation}
\sum_{i}  x_{i} = 2 \xi \, .
\label{Sum}
\end{equation}
The configurations with positive momentum fractions, $x_i\ge 0$, describe the creation of 
quarks, whereas those with  negative momentum fractions, $x_i\le 0$, the 
absorption of antiquarks.

Counting the degrees of freedom fixes the number of independent $P \to \gamma$ TDAs  to 16, 
since each quark, photon and proton have two helicity states (leading to $2^5$ helicity amplitudes) 
and parity relates amplitudes with opposite helicities for all particles. We can equally say that
the photon has spin 1, which  would normally give 24 TDAs as in the $P \to V$ 
where $V$ is a massive vector particle, but gauge invariance provides
us with 8 relations between TDAs, which reduces again the number to 16.

The case $\Delta_T=0$ is simpler since the matrix elements can be written
 only in terms of 4 TDAs. Indeed, since
at $\Delta_T=0$, there is no angular momentum exchanged, the helicity is conserved. We have three
possible processes as $P(\uparrow) \to uud(\uparrow \downarrow \downarrow) +\gamma(\uparrow)$ where
the quark with helicity -1 is either the $u$'s or the $d$, but also
$P(\uparrow) \to uud(\uparrow \uparrow \uparrow) +\gamma(\downarrow)$. Therefore taking this limit
on the complete set of the 16 TDAs should reduce it to 4.

In order to build leading-twist structures (maximising the power of $P^+$), we have 
first to separate the spinor $N(p_1)$ in its small ($N^-\sim \sqrt{1/P^+}$) and large 
($N^+\sim \sqrt{P^+}$) component:
\eqsa{N=(\ks n \ks p + \ks p \ks n) N = N^-+N^+.}
Using the Dirac equation ${\ks p}_1 N(p_1)= M N(p_1)$ and~\ce{eq:decomp_moment}, it is easy to see that 
\eqsa{
&&\ks p N= \frac{M}{2(1+ \xi)}N^+ +{\cal O}(1/P^+) ~\hbox{and}~
\ks n N= \frac{1+ \xi}{2M}N^- +{\cal O}(1/P^+).
}

We then proceed in the 
following way: 
\begin{enumerate}
\item the structures are to  be linear in the photon polarisation vector (through
  scalar products with the momenta ($n$, $p$ and $\Delta_T$), $\gmu$ or $\sigma^{\mu\nu}$).
\item we force the presence of $p$ ($\simeq P$) to help the twist counting in powers of 
$P^+$(therefore the different
leading-twist structures will scale like $(P^+)^{3/2}$);
\item $p$ does not appear in $\ks p N$  since this would remove one power of $P^+$ ;
\item  $p$ does not appear in any scalar products $p.n$, $p.\Delta_T$ and $p.\ep$ 
which would also destroy 
one power of $P^+$;
\item $p$  then only appears inside the parenthesis $(\cdot)_{\alpha\beta}$;
\item we impose the independence of the factors in  $(\cdot)_{\alpha\beta}$ from two different structures; this can be checked by taking the trace
of the product of two structures, and is therefore insured by choosing only independent Fierz (or Dirac) structures  
$\ga$, $\gmu$, $\ga\gmu$, $\sigma^{\mu \nu}$. 
\item Finally, to what concerns the spinor, it has
only two large components. Hence, after a given $(\cdot)_{\alpha\beta}$, 
it  appears
only twice with a different Dirac structure (\eg~$N$ and $\ks \ep N$). 
\end{enumerate}

This construction leads to  define 24 possible independent structures for the transition
proton to vector (whose factors $V_i$, $A_i$ and $T_i$ are  
dimensionless and real function of the momentum fractions $x_i$, 
$\xi$ and $\Delta^2$):
\eqsa{  \label{eq:tda-pgamma}
\lefteqn{4 {\cal F}\Big(\langle     V(p_V)|\, \epsilon_{ijk}u^{i}_{\alpha}(z_1 n) 
u^{j}_{\beta}(z_2 n)d^{k}_{\gamma}(z_3 n)\Big)
\,|P(p_1,s_1) \rangle   
 =M \times}\\ \nn
\lefteqn{\Big(V_1^\ep (\ks p C)_{\alpha \beta } (\ks\ep N^+)_{\gamma}+
M^{-1} V^T_1 (\ep.\Delta_T) (\ks p C)_{\alpha \beta } (N^+)_{\gamma}+
M V^n_1  (\ep.n) (\ks p C)_{\alpha \beta } (N^+)_{\gamma}+ }
\\ \nn
\lefteqn{M^{-1} V_2^{\ep}(\ks p C)_{\alpha \beta } (\sigma^{\Delta_T \ep}  N^+)_{\gamma}+
M^{-2} V^T_2  (\ep.\Delta_T) (\ks p C)_{\alpha \beta } (\ks \Delta\!_T N^+)_{\gamma}+
V^n_2  (\ep.n) (\ks p C)_{\alpha \beta } (\ks \Delta\!_T N^+)_{\gamma}+}
\\\nn
\lefteqn{A_1^\ep(\ks p \ga C)_{\alpha \beta } (\ga \ks\ep  N^+)_{\gamma}+
M^{-1} A^T_1 (\ep.\Delta_T) (\ks p \ga C)_{\alpha \beta } ( \ga N^+)_{\gamma}+
M A^n_1  (\ep.n) (\ks p \ga C)_{\alpha \beta } ( \ga N^+)_{\gamma}+}
\\\nn
\lefteqn{M^{-1} A_2^{\ep}(\ks p \ga C)_{\alpha \beta } (\ga  \sigma^{\Delta_T \ep}  N^+)_{\gamma}+
M^{-2} A^T_2  (\ep.\Delta_T) (\ks p \ga C)_{\alpha \beta } (\ga\ks \Delta\!_T  N^+)_{\gamma} +
}
\\\nn
\lefteqn{
A^n_2  (\ep.n) (\ks p \ga C)_{\alpha \beta } ( \ga  \ks \Delta\!_T N^+)_{\gamma} +
T_1^\ep (\sigma_{p\mu}C)_{\alpha \beta }(\sigma^{\mu\ep}N^+)_{\gamma}+
M^{-1} T^T_1  (\ep.\Delta_T) (\sigma_{p \mu}C)_{\alpha \beta } (\gmu N^+)_{\gamma} +
}
\\\nn
\lefteqn{
M T^n_1  (\ep.n) (\sigma_{p\mu}C)_{\alpha \beta } (\gmu N^+)_{\gamma} +
T_2^\ep (\sigma_{p\ep}C)_{\alpha \beta } (N^+)_{\gamma}+
M^{-2} T^T_2  (\ep.\Delta_T)(\sigma_{p \mu} C)_{\alpha \beta }  (\sigma^{\mu \Delta_T} N^+)_{\gamma}+
}
\\\nn
\lefteqn{
T^n_2  (\ep.n)(\sigma_{p\mu} C)_{\alpha \beta }  (\sigma^{\mu \Delta_T} N^+)_{\gamma}+
M^{-1} T_3^\ep (\sigma_{p\Delta_T }C)_{\alpha \beta } (\ks \ep N^+)_{\gamma}+
}
\\\nn
\lefteqn{
M^{-2} T^T_3  (\ep.\Delta_T) (\sigma_{p \Delta_T}C)_{\alpha \beta } (N^+)_{\gamma}
+T^n_3  (\ep.n) (\sigma_{p\Delta_T}C)_{\alpha \beta } (N^+)_{\gamma}+
M^{-1} T_4^\ep (\sigma_{p\ep }C)_{\alpha \beta } (\ks \Delta\!_T N^+)_{\gamma} +
}
\\\nn
\lefteqn{
M^{-3} T^T_4 (\ep.\Delta_T) (\sigma_{p \Delta_T}C)_{\alpha \beta } (\ks \Delta\!_T N^+)_{\gamma}+
M^{-1} T^n_4  (\ep.n) (\sigma_{p\Delta_T}C)_{\alpha \beta } (\ks \Delta\!_T N^+)_{\gamma}\Big),}
}
where $\sigma^{\mu \nu}\equiv \frac{1}{2}[\gmu,\gnu]$ and  $C$ is the charge-conjugation matrix.

The quark fields in the matrix element of~\ce{eq:tda-pgamma} are defined according to the prescription
of Mandelstam~\cite{Mandelstam:1962mi} in order to make the latter QED gauge invariant. In the proton
to photon case, gauge invariance of the r.h.s of~\ce{eq:tda-pgamma} implies that the latter vanishes when $\ep(p_\gamma)$ is replaced by $p_\gamma$.

At the leading-twist accuracy, this provides us with  8 relations                               :
\eqsa{ \label{eq:GI-tdapgamma}
&V^\ep_1  (1-\xi) \frac{M}{2(1+\xi)}+V_1^T \frac{\Delta_T^2}{M}+V^n_1 \frac{(1-\xi)M}{2}=
V^\ep_1+\frac{V^\ep_2}{2M}  (1-\xi) \frac{M}{2(1+\xi)}+V_2^T \frac{\Delta_T^2}{M^2}+V^n_2 
\frac{1-\xi}{2}=0,\nn\\
&A^\ep_1  (1-\xi) \frac{M}{2(1+\xi)}+A_1^T \frac{\Delta_T^2}{M}+A^n_1 \frac{(1-\xi)M}{2}=
A^\ep_1+\frac{A^\ep_2}{2M}  (1-\xi) \frac{M}{2(1+\xi)}+A_2^T \frac{\Delta_T^2}{M^2}+A^n_2 \frac{1-\xi}{2}=0,\nn\\
&\frac{T^\ep_1}{2}  (1-\xi) \frac{M}{2(1+\xi)}+T_1^T \frac{\Delta_T^2}{M}+T^n_1 \frac{(1-\xi)M}{2}=
T^\ep_2+\frac{T^\ep_3}{M}  (1-\xi) \frac{M}{2(1+\xi)}+T_3^T \frac{\Delta_T^2}{M^2}+T^n_3 \frac{1-\xi}{2}=0,\nn\\
&~~~~T^\ep_1+T_2^T \frac{\Delta_T^2}{M^2}+T^n_2 \frac{1-\xi}{2}  =
T^\ep_3+T^\ep_4+T_4^T \frac{\Delta_T^2}{M^2}+T^n_4 \frac{1-\xi}{2}  =0.\nn\\
}

This effectively reduces the number of $P \to \gamma$ TDAs  to 
16 as expected from the number of helicity amplitudes for the process $P \to q q q \gamma$
, and we have 

\eqsa{ 
\lefteqn{4 {\cal F}\Big(\langle     \gamma(p_\gamma)|\, \epsilon_{ijk}u^{i}_{\alpha}(z_1 n) 
u^{j}_{\beta}(z_2 n)d^{k}_{\gamma}(z_3 n)
\,|p(p_1,s_1) \rangle   \Big)
 =M \times}
\nn\\ 
\lefteqn{\Big(V_1^\ep(x_i,\xi,\Delta^2) (\ks p C)_{\alpha \beta} 
[
(\ks\ep N^+)_{\gamma} 
-\frac{M}{1+\xi}(\ep.n)(N^+)_{\gamma}
- \frac{2(\ep.n)}{1-\xi}(\ks \Delta\!_T N^+)_{\gamma}
]
+}\nn\\
\lefteqn{~\frac{V^T_1(x_i,\xi,\Delta^2)}{M}
[(\ep.\Delta_T) -\frac{2 \Delta^2_T}{1-\xi}(\ep.n)](\ks p C)_{\alpha \beta } (N^+)_{\gamma}+
}\nn\\
\lefteqn{~\frac{V_2^{\ep}(x_i,\xi,\Delta^2)}{M} (\ks p C)_{\alpha \beta } 
[(\sigma^{\Delta_T \ep}  N^+)_{\gamma}-\frac{M (\ep.n)}{2(1+\xi)}(\ks \Delta\!_T N^+)_{\gamma}
]
+}\nn\\
\lefteqn{~\frac{V^T_2(x_i,\xi,\Delta^2)}{M^2}[(\ep.\Delta_T) -\frac{2 \Delta^2_T}{1-\xi}(\ep.n)] (\ks p C)_{\alpha \beta } 
(\ks \Delta\!_T N^+)_{\gamma}+
}\nn\\
\lefteqn{~A_1^\ep(x_i,\xi,\Delta^2) (\ks p \ga C)_{\alpha \beta} 
[
(\ga \ks\ep N^+)_{\gamma} 
-\frac{M}{1+\xi}(\ep.n)(\ga N^+)_{\gamma}
- \frac{2(\ep.n)}{1-\xi}(\ga \ks \Delta\!_T N^+)_{\gamma}
]
+}\nn\\
\lefteqn{~\frac{A^T_1(x_i,\xi,\Delta^2)}{M}
[(\ep.\Delta_T) -\frac{2 \Delta^2_T}{1-\xi}(\ep.n)](\ks p \ga C)_{\alpha \beta } (\ga N^+)_{\gamma}+
}\nn\\
\lefteqn{~\frac{A_2^{\ep}(x_i,\xi,\Delta^2)}{M} (\ks p \ga C)_{\alpha \beta } 
[(\ga \sigma^{\Delta_T \ep}  N^+)_{\gamma}-\frac{M (\ep.n)}{2(1+\xi)}(\ga \ks \Delta\!_T N^+)_{\gamma}
]
+}\nn\\
\lefteqn{~\frac{A^T_2(x_i,\xi,\Delta^2)}{M^2}[(\ep.\Delta_T) -\frac{2 \Delta^2_T}{1-\xi}(\ep.n)] (\ks p \ga C)_{\alpha \beta } 
(\ga \ks \Delta\!_T N^+)_{\gamma}+
}\nn\\
\lefteqn{~T_1^\ep(x_i,\xi,\Delta^2) (\sigma_{p \mu}C)_{\alpha \beta }
[(\sigma^{\mu\ep}N^+)_{\gamma}
-\frac{M (\ep.n)}{2(1+\xi)}(\gmu N^+)_{\gamma}
-\frac{2 (\ep.n)}{(1-\xi)}(\sigma^{\mu \Delta_T} N^+)_{\gamma}
]+
}\nn\\
\lefteqn{~\frac{T^T_1(x_i,\xi,\Delta^2)}{M} 
[(\ep.\Delta_T) -\frac{2 \Delta^2_T}{1-\xi}(\ep.n)]
(\sigma_{p \mu}C)_{\alpha \beta }(\gmu N^+)_{\gamma}+
}\nn\\
\lefteqn{~T_2^\ep(x_i,\xi,\Delta^2)
[(\sigma_{p\ep}C)_{\alpha \beta } 
-\frac{2(\ep.n)}{(1-\xi)}(\sigma_{p \Delta_T}C)_{\alpha \beta }
]
(N^+)_{\gamma}+
}\nn\\
\lefteqn{~\frac{T^T_2(x_i,\xi,\Delta^2)}{M^2}[(\ep.\Delta_T) -\frac{2 \Delta^2_T}{1-\xi}(\ep.n)]
(\sigma_{p \mu} C)_{\alpha \beta }  (\sigma^{\mu \Delta_T} N^+)_{\gamma}+
}\nn\\
\lefteqn{~\frac{T_3^\ep(x_i,\xi,\Delta^2)}{M}
(\sigma_{p\Delta_T }C)_{\alpha \beta } 
[(\ks \ep N^+)_{\gamma}-\frac{M (\ep.n)}{(1+\xi)}(N^+)_{\gamma}
-\frac{2(\ep.n)}{(1-\xi)}(\ks \Delta_T N^+)_{\gamma}]+
}\nn\\
\lefteqn{~\frac{T^T_3(x_i,\xi,\Delta^2)}{M^2}[(\ep.\Delta_T) -\frac{2 \Delta^2_T}{1-\xi}(\ep.n)]
(\sigma_{p \Delta_T}C)_{\alpha \beta } (N^+)_{\gamma}+
}\nn\\
\lefteqn{~\frac{T_4^\ep(x_i,\xi,\Delta^2)}{M}
[(\sigma_{p\ep }C)_{\alpha \beta } 
-\frac{2(\ep.n)}{1-\xi} (\sigma_{p\Delta_T}C)_{\alpha \beta }
]
(\ks \Delta\!_T N^+)_{\gamma}+
}\nn \\
\lefteqn{~\frac{T^T_4(x_i,\xi,\Delta^2)}{M^3} [(\ep.\Delta_T) -\frac{2 \Delta^2_T}{1-\xi}(\ep.n)]
(\sigma_{p \Delta_T}C)_{\alpha \beta } (\ks \Delta\!_T N^+)_{\gamma}\Big).
\label{eq:tda-pgammaGI}
}}

As discussed earlier, the $\Delta_T=0$ case is much simpler since it involves only 
4 TDAs to describe the proton to photon transition. Moreover, in the Bjorken scaling
which interests us, $\Delta_T$ is in any case supposed to be small, making this limit
$\Delta_T=0$ particularly fruitful to consider.

The four expected TDAs for  $p \to \gamma$ TDAs  are straightforwardly obtained from~\ce{eq:tda-pgammaGI}
by setting $\Delta_T=0$ :
\eqsa{ \label{eq:tda-pgammaGI-dt0}
\lefteqn{4  {\cal F} \Big( \langle    \gamma(p_\gamma)|\, 
\epsilon_{ijk}u^{i}_{\alpha}(z_1 n) u^{j}_{\beta}(z_2 n)d^{k}_{\gamma}(z_3 n)
\,|P(p_1,s_1) \rangle \Big)
 =M \times}\\ 
 \lefteqn{\Big(V^\ep_1(x_i,\xi,\Delta^2)\ (\ks p C)_{\alpha \beta }  \Big[ 
(\ks\ep N^+)_{\gamma}-\frac{M}{1+\xi}(\ep.n)  (N^+)_{\gamma}\Big] +}\nn\\
 \lefteqn{
~A^\ep_1(x_i,\xi,\Delta^2)\ (\ks p \ga C)_{\alpha \beta } \Big[ 
(\ga \ks\ep  N^+)_{\gamma}-\frac{M}{1+\xi}(\ep.n)( \ga N^+)_{\gamma}\Big]
+}\nn\\
 \lefteqn{~T^\ep_1(x_i,\xi,\Delta^2)\  (\sigma_{\mu p}C)_{\alpha \beta }  \Big[(\sigma^{\mu\ep}N^+)_{\gamma}
-\frac{M}{1+\xi}\frac{(\ep.n)}{2}(\gmu N^+)_{\gamma} \Big]+
T^\ep_5(x_i,\xi,t)\ (\sigma_{p \ep}C)_{\alpha \beta } (N^+)_{\gamma}\Big).}\nn
}

\section{Amplitude calculation at $\Delta_T=0$ and model-independent predictions}

Let us now consider the calculation of the helicity amplitude in the $\Delta_T=0$ limit. 
At  leading order in 
$\alpha_{S}$, the helicity amplitude  ${\cal M}_{\lambda_1,\lambda_2,s_1,s_2} $ for the reaction 
\begin{equation}
~~~~~~~~~~~~~~~~~~~~~~~~~~~~~~~\gamma^\star(q,\lambda_1) ~ P(p_1, s_1) \to  P(p_2,s_2) ~\gamma(p_\gamma,\lambda_2) 
\end{equation}
 is  calculated similarly to the baryonic form-factor \cite{ERBL,CZ}. It reads

\begin{eqnarray}\label{hardamp}
{\cal M}_{\lambda_1,\lambda_2,s_1,s_2}
 &\propto&  e \bar u(p_2,s_2) 
\ks \ep^{\lambda_1} \ks \ep^{\lambda_2} \ga u(p_1,s_1)
 M \frac{(\alpha_S(Q^2))^2}{Q^4} \times \nn\\&&
\int\limits_{1+\xi}^{-1+\xi} d^3x_i \int\limits_0^1 d^3y_j 
\delta(\sum x_i -2\xi)\delta (\sum y_j -1)
\sum\limits_{\alpha=1}^{10} T_{\alpha}(x_{i},y_{j},\xi,\Delta^2)\nn\\
\end{eqnarray}
where the coefficients\footnote{whose details (omitted due to lack of space)
will be presented in a forthcoming publication~\cite{bdVCS}.} $T_{\alpha}$ include both
the proton to photon TDAs and the final-state proton DAs.
The  structure $\bar u(p_2,s_2) \ks \ep^{\lambda_1} \ks \ep^{\lambda_2} \ga u(p_1,s_1)$ selects 
opposite  helicity states for the final and initial protons. The same statement holds for the photons. 
This is a model independent result at $\Delta_T=0$ as well as the scaling $\frac{(\alpha_S(Q^2))^2}{Q^4}$
up to logarithmic corrections due to the evolution of the TDAs and DAs.

\section{Conclusions and perspectives}

We have defined the 16 proton to photon Transition Distribution Amplitudes 
entering the description of backward virtual Compton scattering on proton target. 
Since the study in terms of GPDs of the 
latter process in the forward region has been very fruitful to understand the underlying structure
of the hadron, we foresee that the corresponding one with TDAs of the backward region be of equal 
importance, if not more since it involves the exchange of 3 quarks.

We have also calculated the amplitude for the process $\gamma^* P \to P' \gamma $ 
in terms of the TDAs. 
 In order to provide with theoretical evaluations
of cross sections, we still have to develop an adequate model for the TDAs $V_i$, $A_i$ and $T_i$. 
This may be done 
through the introduction of quadruple distributions, which generalise the double distributions 
introduced by Radyushkin~\cite{Radyushkin} in the GPD case. Similarly to this latter case, 
it will also 
ensure the proper polynomiality and support properties of the TDAs.
A limiting value of the TDA for $\xi \to 1$ may be derived by considering the soft photon limit 
of the scattering amplitude and may be used as a model input in these quadruple distributions, whereas
for the GPDs the diagonal limit, \ie~the parton distribution functions, was used as input.

Model independent predictions follow from the way we propose to factorise the amplitude :
only helicity amplitudes with opposite signs for both protons 
and photons will be nonzero at $\Delta_T=0$. 
Furthermore, the amplitude scales as $\frac{(\alpha_S(Q^2))^2}{Q^4}$ as do the similar amplitudes 
for backward electroproduction of mesons, \ie~$\gamma^\star P \to P' \pi$  or  
$\gamma^\star P  \to P' \rho$. Observation of such a universal scaling law
would provide with indications that the  picture holds and dominates over a purely hadronic 
model as considered in~\cite{Laveissiere:2004gr}, where data are presented for low energies but 
for $Q^2 = 1 $GeV$^2$.

Let us finally stress that, first, in the backward region considered here, there is  
almost no Bethe-Heitler contribution: the experimentally measured cross sections will 
depend bilinearly on the TDAs; secondly, the same matrix elements of \ce{eq:tda-pgammaGI}  
factorise in the amplitude $ P \bar P \to \gamma^\star \gamma$, which 
may be studied at GSI. This universality makes the TDAs an essential tool for the 
generalisation of the hadronic studies carried at electron machines to complementary 
studies to be carried in proton antiproton experiments. Thus, the properties of these TDAs 
are planned to be studied both at the upgraded JLab experiments and with PANDA and 
PAX~\cite{Panda} at GSI .

\subsection*{Acknowledgments.}
 
We are thankful to P. Bertin, V. Braun, M. Gar\c con, C.D. Hyde-Wright, A.V. Radyushkin 
F. Sabati\'e for useful discussions.
This work is partly supported by the French-Polish scientific agreement Polonium, 
the Polish Grant 1 P03B 028 28, the
Fonds National de la Recherche Scientifique (FNRS, Belgium), the ECO-NET program, contract 
12584QK and the Joint Research Activity "Generalised Parton Distributions" of the european I3 program
Hadronic Physics, contract RII3-CT-2004-506078. We finally thank the organisers of the Vth 
Conference on Hadronic Physics  at Trieste for their kind invitation.

\end{document}